\documentclass[9pt,amsmath,amssymb,nofootinbib]{revtex4}
\usepackage{dcolumn}
\usepackage{bm}
\usepackage[all, knot]{xy}
\xyoption{arc}
\begin{document}
\title{Fock space, quantum fields and $\kappa$-Poincar\'e symmetries}
\author{Michele Arzano}
\email{marzano@perimeterinstitute.ca}
\affiliation{Perimeter Institute for Theoretical Physics\\
31 Caroline St. N, N2L 2Y5, Waterloo ON, Canada}
\author{Antonino Marcian\`o}
\email{antonino.marciano@roma1.infn.it}
\affiliation{Dipartimento di Fisica\phantom{a}and INFN\\
Universit\`a degi Studi di Roma ``La Sapienza" \\
P.le A. Moro 2, 00185 Roma, Italy}
\begin{abstract}
\begin{center}
{\bf Abstract}
\end{center}
We study the quantization of a linear scalar field, whose symmetries are described by the $\kappa$-Poincar\'e Hopf-algebra, via deformed Fock space construction. The one-particle sector of the theory exhibits a natural (planckian) cut-off for the field modes. At the multi-particle level the non-trivial co-algebra structure of $\kappa$-Poincar\'e leads to a deformed bosonization in the construction of Fock space states.  These physical states carry energy-momentum charges which are divergenceless and obey a deformed dispersion relation.
 \end{abstract}
\maketitle
\section{Introduction}
Local quantum fields on Minkowski space-time provide a successful framework for an effective description of particle physics up to the energies probed by current experiments.  The Planck energy ($E_p\sim10^{28}eV$) sets the scale at which such description is believed to break down.  Indeed at planckian regimes the field quanta would probe the background space-time at scales for which quantum-gravitational effects become important and new physics described by a general theory of quantum gravity would take the stage.\\  Quantum field theories on non-commutative spaces are an example of generalizations of ordinary quantum field theory which find their motivation from certain quantum gravity scenarios and whose new features are expected to emerge in the ultraviolet \cite{Doplicher:1994tu, Douglas:2001ba, Szabo:2001kg, AmelinoCamelia:2002mu}.  The view that such theories might enjoy ``deformed" relativistic symmetries (rather than breaking standard Poincar\'e invariance through the emergence of the non-commutativity scale) has become increasingly popular in recent years \cite{Chaichian:2004za, Chaichian:2004yh}.  Such deformed symmetries are generally described by a ``quantization" of standard relativistic symmetries in terms of quantum groups (non-cocommutative Hopf algebras) \cite{k-literature}. 
In the present work we focus our attention on the $\kappa$-Poincar\'e Hopf algebra \cite{Lukierski:1992dt, Majid-Ruegg} as one candidate for quantum group description of relativistic symmetries at the Planck scale \cite{Amelino_fate}.  The deep connection between $\kappa$-Poincar\'e and $\kappa$-Minkowski space-time, a ``Lie algebra"-type of non-commutative space-time, has been known for quite a while \cite{Majid-Ruegg}.  Indeed the $\kappa$-Poincar\'e Hopf algebra can be used to describe the symmetries of such non-commutative space-time \cite{Majid-Ruegg, Ale2}.    Much of the recent interest towards these type of deformed symmetries has been motivated by various arguments which suggest that $\kappa$-Poincar\'e-type of quantum group symmetries might emerge in the low-energy limit of three (and possibly four) dimensional quantum gravity \cite{qg3D, Amelino-Camelia:2003xp, Freidel:2005me} with the dimensionful deformation parameter $\kappa$ set by the Planck energy.  Moreover earlier interest in the study of scenarios with $\kappa$-Poincar\'e symmetries arose from the possibility of detecting experimental signatures of the deformed kinematics associated with such symmetries \cite{pheno}.\\
Classical field theories with $\kappa$-type of deformed symmetries have now been studied extensively \cite{Ale2, knoeth, Freikono, arma, Meljanac:2007xb, FreiKoNo2, Dimitrijevic:2003wv} however much work seems to be needed to reach a satisfactory understanding of their quantum counterparts.  One of the early proposals for the quantization of a self-interacting scalar field, by one of us and Amelino-Camelia \cite{arzame}, was based on a generalization of the path integral quantization method which in a $\kappa$-deformed context seemed more straightforward than the canonical approach (see section IV below for a review of some recent attempts on the quantization of fields with $\kappa$-Poincar\'e symmetries).  Recent works on the derivation of Noether charges \cite{knoeth, arma} associated with $\kappa$-deformed symmetries of classical fields have prompted new attention to the problem of canonical quantization.\\ 
In the present paper we discuss a new strategy for the quantization of a free field theory with space-time symmetries described by the $\kappa$-Poincar\'e Hopf algebra.  We start from the construction of a one-particle Hilbert space, we define an appropriate Fock space, creation and annihilation operators associated to a given basis of modes, and we study the properties of energy and momentum charges carried by the quantum states of the theory.  The very first step of our approach to quantization will be the introduction of a (positive definite) inner product on the space of solutions of the deformed equation of motion.  To this end we borrow the results of \cite{arma} in which such inner product was derived using the tools of symplectic geometry of classical fields.  Thus the present analysis can be seen as a natural continuation of the work started in \cite{arma}.\\ 
Below is the plan of the paper.  In the following section we review the basic properties of the $\kappa$-Poincar\'e Hopf algebra in the so-called bicrossproduct basis \cite{Majid-Ruegg} and recall some results from our previous work \cite{arma} which will turn out to be useful in our construction.  In Section III we focus on the quantization of free fields.  In the first subsection we start with an overview of the usual Hilbert (and Fock) space construction in the presence of standard space-time symmetries.  We then proceed, in Section III.B, with the definition of a one-particle Hilbert space in the context of deformed symmetries described by the $\kappa$-Poincar\'e Hopf algebra and show how a natural cut-off, set by the deformation parameter $\kappa$, arises in the construction.  In the Section III.C we use the one-particle Hilbert space defined in the previous section and introduce a ``deformed bosonization" in order to build the full Fock space of the theory.  Then in the following Section III.D we study the energy and momentum charges carried by the states of our deformed theory.  In Section IV we review and comment on some recent works related to our results.  Section V contains closing remarks and some suggestions for future studies.

\section{Preliminaries}
\subsection{$\kappa$-Poincar\'e Hopf algebra}
In the following we will consider $\kappa$-deformations of relativistic symmetries described by a Hopf algebra in which both the co-algebra sector and the product (commutator) of the algebra elements (symmetries generators) are deformed.  In particular we will focus on the example of the $\kappa$-Poincar\'e Hopf algebra $\mathcal{P}_{\kappa}$ \cite{Lukierski:1992dt} in the {\it bicrossproduct basis} \cite{Majid-Ruegg}.\\ 
It is well known that the standard Poincar\'e algebra $\mathcal{P}$  can be endowed with the structure of a (trivial) Hopf algebra.  If $Y=\{P_{\mu},M_{\mu\nu}\}$ are the generators of $\mathcal{P}$, the (trivial) co-product, co-unit and antipode can be defined respectively as
\begin{equation}
\label{trivialcopro}
\Delta(Y)=Y\otimes 1+1\otimes Y,\,\,\,\,
\epsilon(Y)=0,\,\,\,\,S(Y)=-Y\, .
\end{equation}
The main feature of the $\kappa$-Poincar\'e Hopf algebra $\mathcal{P}_{\kappa}$ is that its coproducts are, in general, non-cocommutative, in particular in the bicrossproduct basis they read
\begin{eqnarray}
\Delta(P_0)&=&P_0\otimes 1+1\otimes P_0\,\,\,\,\,\,\Delta(P_j)=P_j\otimes 1+e^{-P_0/\kappa}\otimes P_j \nonumber \\
\Delta(M_{j})&=&M_{j}\otimes 1+1\otimes M_{j} \nonumber \\
\Delta(N_j)&=&N_j\otimes 1+e^{-P_0/\kappa}\otimes N_j+\frac{\epsilon_{jkl}}{\kappa}P_k\otimes N_l\, . \label{copro}
\end{eqnarray}\\
One also has deformed commutators for the generators of translations $P_0,P_j$, rotations $M_j$ and boosts $N_j$
\begin{eqnarray}
&[P_{0},P_{j}]=0 \qquad [M_j,M_k]=i \epsilon_{jkl}M_l \qquad [M_j,N_k]=i \epsilon_{jkl}N_l  \qquad [N_j,N_k]=i \epsilon_{jkl}M_l  \nonumber\\
&[P_0,N_l]=-iP_l \qquad [P_l,N_j]=-i\delta_{lj}\Big( \frac{\kappa}{2}  \left(1-e^{-\frac{2 P_0}{\kappa}} \right) +\frac{1}{2 \kappa} \vec{P}^2 \Big)+  \frac{i}{\kappa}P_l P_j \nonumber \\
&[P_0,M_k]=0 \qquad [P_j,M_k]=i \epsilon_{jkl}P_l
\end{eqnarray}\\
where $\epsilon_{jkl}$ is the Levi-Civita symbol and  $j,k,l$ are spatial indices.\\
The non-trivial antipodes consistent with the coproducts above are 
\begin{eqnarray}
S(M_l)&=&-M_l \nonumber\\
S(P_0)&=&-P_0 \nonumber\\
S(P_l)&=&-e^{\frac{P_0}{\kappa}}P_l \nonumber\\
S(N_l)&=&-e^{\frac{P_0}{\kappa}}N_l+\frac{1}{\kappa}\epsilon_{ljk}e^{\frac{P_0}{\kappa}}P_j M_k  \label{anti}\, ,
\end{eqnarray}
while for the co-units one has
\begin{equation}
\epsilon(P_{\mu})=\epsilon(M_j)=\epsilon(N_k)=0.
\end{equation}
The (deformed) mass Casimir $C_{\kappa}$ of the $\kappa$-Poincar\'e Hopf algebra links the  generators of translations and is given by 
\begin{equation}
C_{\kappa}=\left( 2\kappa \sinh \left( \frac{P_0}{2 \kappa}\right)  \right)^2-\vec{P}^2e^{\frac{P_0}{\kappa}}.  
\end{equation}
As mentioned in the introduction the $\kappa$-Poincar\'e Hopf algebra describes the symmetries of $\kappa$-Minkowski non-commutative space-time \cite{Ale2}
\begin{equation}
\label{kappaM}
[x_i,t]=\frac{i}{\kappa}x_i\,\,\,\,\,\,\,\,\,\,[x_i,x_j]=0\,\,.
\end{equation}
For further details on $\kappa$-Poincar\'e in the bicrossproduct basis and its relation with $\kappa$-Minkowski noncommutative space-time we refer to the literature (see e.g. \cite{Ale2} and references therein).

\subsection{Symplectic geometry of classical $\kappa$-field theories}
In this section we summarize the key results obtained in \cite{arma} where we proposed a description of classical field theories with Hopf-algebra of space-time symmetries using the tools of symplectic geometry. 
The idea behind a symplectic geometrical formulation of classical field theories is to provide a canonical formulation of the phase space which does not spoil the symmetries of the theory.  Such description turns out to be very useful in the deformed field theory case as it gives a clear identification of the Noether charges associated with space-time symmetries and more importantly, for the purposes of the present work, it provides a natural candidate for an inner product which will lead to the definition of the Hilbert space of our quantum fields.\\
We start with a classical field theory with standard (Poincar\'e) relativistic symmetries.  We will consider a real massless scalar field with Lagrangian
\begin{equation}
L=\int_{M}\frac{1}{2}(\partial_{\alpha}\Phi\partial^{\alpha}\Phi)
\end{equation}
where $M$ is the space-time manifold on which the theory is defined which in the undeformed case is just standard Minkowski space-time. The phase space of such theory is the infinite dimensional manifold $\Gamma$ whose elements are the pairs $\{\Phi,\Pi\}$, where $\Pi=\frac{\partial \mathcal{L}}{\partial \dot{\Phi}}$ is the momentum (density) canonically conjugate to the field $\Phi$.  The definition of $\Pi$ requires singling out a time coordinate $t$, a choice that obscures the relativistic invariance of the theory.  The crucial observation \cite{Crnkovic:1986ex} is that there is a one-to-one correspondence between the points of the classical phase space and the solutions of the Klein-Gordon equation of motion thanks to the uniqueness properties of the initial value problem for such equation in Minkowski space.  This allows one to {\it identify} points in phase space with solutions of the classical equation of motion i.e. the phase space $\Gamma$ {\it is} the space of solutions $\mathcal{S}$.
The next step in the formalism is the introduction of a symplectic structure through the definition of a non-degenerate two-form $\omega$.  To do so one needs to specify the differential structure (functions, tangent ($T\mathcal{S}$) and cotangent ($T^* \mathcal{S}$) spaces) on $\mathcal{S}$.  For details we refer to \cite{arma} and references therein.  The non-degenerate, symplectic 2-form $\omega$ one needs is given by 
\begin{equation}
\label{omega1}
\omega\equiv\delta\left(\int_{\Sigma}d\sigma_{\alpha} J^{\alpha} \right)
\end{equation}
with the ``symplectic potential current" $J^{\alpha}$ defined by
\begin{equation}
J^{\alpha}\equiv\frac{\partial L}{\partial (\Phi _{,\alpha})}\delta \Phi(x)
\end{equation}
where $\delta \Phi(x)\in T^*_{\Phi}\mathcal{S}$ and $\delta$ is the exterior derivative.  It can be easily shown \cite{Crnkovic:1986ex} that $\omega$ is Poincar\'e invariant and for a vector field tangent to the orbit of space-time symmetry $V_s\in T\mathcal{S}$ 
\begin{equation}
\omega(V_s)=-\delta (Q_s)
\end{equation}
where $Q_s$ is the generator of the symmetry.  In particular for a choice of $\Sigma$ as the standard $t=0$ space-like hypersurface (\ref{omega1}) reduces to the more familiar expression for the symplectic form on the phase space manifold
\begin{equation}
\omega=\frac{1}{2}\int_{\Sigma_t}\delta\Pi \wedge \delta\Phi
\end{equation}
and
\begin{equation}
\omega(V_s)=\frac{1}{2}\int_{\Sigma_t}\left(\delta\Pi(V_s) \delta\Phi-\delta\Pi \delta\Phi(V_s)\right)\, .
\end{equation}
$\mathcal{S}$, as a vector space, is isomorphic to the tangent space at any given point $T_{\Phi}\mathcal{S}$ and through this correspondence the 2-form $\omega$ and the one form $\omega(V_s)$ can be used to define, respectively, a bilinear and a linear functional on $\mathcal{S}$.  The bilinear will induce a symplectic product on $\mathcal{S}$ 
\begin{equation}
\omega(\Phi_1,\Phi_2)=\frac{1}{2}\int_{\Sigma_t}(\Pi_1\Phi_2-\Phi_1\Pi_2)
\end{equation}
and the linear functional $\omega(V_s)$ evaluated on a solution $\Phi$ will give the (conserved) value of the Noether charge associated with the symmetry
\begin{equation}
\omega(V_s)(\Phi)=\frac{1}{2}\int_{\Sigma_t}\left((\mathcal{L}_{(V_s)}\Pi) \Phi - \Pi (\mathcal{L}_{(V_s)}\Phi) \right)=-Q_s\, 
\end{equation}
where the contraction of a 1-form with the Killing vector field $V_s$ is replaced by the Lie derivative\footnote{From the expression of the Lie derivative of a general p-form $\mathcal{L}_{V_s}F=(\delta F)(V_s)+\delta(F(V_s))$ which for a function (0-form) implies $\mathcal{L}_{V_s}f=(\delta f)(V_s)$.} of $\Phi\in\mathcal{S}$ and its canonically conjugate momentum $\Pi$.
When the Killing vector fields correspond to (external) space-time symmetries, $Y\in\mathcal{P}$, we will use the following notation for the action of the Lie algebra element associated with the symmetry on functions
\begin{equation}
\mathcal{L}_{Y}\Phi=Y\vartriangleright \Phi\, 
\end{equation}
to stress the fact that $\mathcal{S}$ is an $\infty$-dimensional representation of the Poincar\'e algebra.\\
On the space of complex solutions of the (massless) Klein-Gordon equation, $\mathcal{S}^{\mathbb{C}}$, our symplectic structure will define an hermitian inner product 
\begin{equation}
(\Phi_1,\Phi_2)=-2i\,\omega(\Phi^*_1,\,\Phi_2)
\end{equation}
where $\Phi^*_i$ denotes the complex conjugate of $\Phi\in\mathcal{S}^{\mathbb{C}}$.  It will be useful for our purposes to write such product in an explicit covariant form as 
\begin{equation}
\label{stinnerp}
(\Phi_1,\Phi_2)=\int\frac{d^4 p}{(2\pi)^3}\delta(\mathcal{C}(p))\frac{p_0}{|p_0|}\tilde{\Phi^*}_1{(-p)}\tilde{\Phi}_2(p)
\end{equation}
where $\tilde{\Phi}_i(p)$, $\tilde{\Phi^*}_i(p)$ are the Fourier transforms of $\Phi_i$, $\Phi^*_i$ and $C(p)$ is the relativistic ``mass" Casimir which sets the Fourier parameters $p_{\mu}$ on-shell according to the equation of motion.\\
Noether charges associated with space-time translations can be expressed in terms of the action of the symmetry generators $P_{\mu}$ acting on the fields $\Phi$ using the conserved symplectic product
\begin{equation}
Q_{\mu}=\frac{1}{2}(\Phi, P_{\mu} \vartriangleright \Phi)\, . 
\end{equation}\\
We now turn our attention to the deformed case.  In the standard case plane waves on-shell are solutions to the Klein-Gordon equation and provide a basis for the standard space of solutions.
In the $\kappa$-Poincar\'e Hopf algebra symmetry case plane waves will still be solutions of the deformed Klein-Gordon equation \cite{kDirac}
\begin{equation}
\label{kappakg}
\mathcal{C}_{\kappa}(P)\,\Phi=0\,.
\end{equation}  
However wave exponentials labeled by a given value of the spatial momentum will combine according to the coproduct structure of the non-trivial co-algebra sector of the Hopf algebra. Seen from a dual space-time point of view \cite{arzame} this corresponds to introducing a $*$-multiplication for the algebra of functions on space-time coordinates.  
Roughly speaking any such $*$-structure corresponds to a different choice of normal ordering for functions of non-commuting coordinates or to a different choice of basis for the generators of the $\kappa$-Poincar\'e (Hopf) algebra \cite{Ale2}.  Unlike the standard Lie-algebra symmetry case, in $\kappa$-Poincar\'e in the bicrossproduct basis, for each mode $\vec{p}$ the positive and negative roots of the deformed mass Casimir $\mathcal{C}_{\kappa}(p)$ will not be equal in modulus.  Infact there will be two roots\footnote{$\mathcal{C}_{\kappa}(p)=0$ has a countable infinite set of complex roots.  As it will become clear from our considerations below, each pair of such complex roots will correspond to a copy of the same deformed space of solutions.  For our purposes the restriction to a single pair of roots, which in the limit $\kappa\rightarrow\infty$ reproduces the standard positive and negative roots, will be enough to construct a model of field theory with {\it deformed} relativistic symmetries.}:
\begin{equation}
\label{kroots}
\omega^{\pm}(\vec{p})=\kappa\log(\Omega(\vec{p})_{\pm})\, ,
\end{equation}
where $\Omega_{\pm}(\vec{p})$ is a  function of $|\vec{p}|$ and in our specific case reads
\begin{equation}
\Omega(\vec{p})_{\pm}=\frac{1}{1\mp\frac{|\vec{p}|}{\kappa}}\, .
\end{equation}
Notice how the positive root becomes complex for transplanckian ($|\vec{p}|>\kappa$) modes.
In the limit $\kappa\rightarrow\infty$ each solution approaches the value of the standard roots ${\pm}\omega =\pm|\vec{p}|$.\\
We denote with  $\mathcal{S}^{\mathbb{C}}_{\kappa}$ the spaces of (complex) solutions of the $\kappa$-Klein-Gordon equation (\ref{kappakg}).
The set of plane waves $\left\{\phi_{\vec{p} \pm}^{\kappa}\right\}$, with the subscript $\vec{p} \pm$ specifying plane waves whose momentum labels are on the positive or negative deformed mass-shell, provides a basis for the associated vector space $\mathcal{S}^{\mathbb{C}}_{\kappa}$. This can be easily seen \cite{arma} introducing a linear map 
\begin{equation}
\label{map}
\mathfrak{m}:\mathcal{S}^{\mathbb{C}}_{\kappa} \rightarrow \mathcal{S}^{\mathbb{C}} 
\end{equation}
which associates every element of $\mathcal{S}^{\mathbb{C}}_{\kappa}$ to its ``classical" counterpart in $\mathcal{S}^{\mathbb{C}}$.
Positive and negative energy solutions are related by {\it deformed} complex conjugation which on plane waves acts as follows 
\begin{equation}
\bar{\phi}_{\vec{p}_+}^{\kappa}=\phi_{\dot{-}\vec{p}_+}^{\kappa}=\phi_{\vec{k}_-}^{\kappa}
\end{equation}
with $\vec{k}$ given by $\vec{k}=-\vec{p}\,e^{\frac{\omega^+}{\kappa}}$ and the label $\dot{-}\vec{p}_{\pm}$ is a short hand notation meaning that we are taking the antipode (\ref{anti}) of a four-momentum on-shell in the plane wave.\\
Both in the standard and in the $\kappa$-deformed case the spaces of positive and negative energy solutions can be shown to be isomorphic to the spaces of functions on the positive and negative mass-shell, respectively, through (non-commutative) Fourier transforms \cite{Ale2}.  As a vector space, $\mathcal{S}^{\mathbb{C}}_{\kappa}$ can be endowed with a symplectic structure if we require  $\mathfrak{m}$ to be a {\it Poisson map} (see \cite{arma}).  Such symplectic structure defines a bilinear functional $\omega_{\kappa}(\Phi^{\kappa}_1,\Phi^{\kappa}_2)$ from which one can define the inner product\footnote{Here we drop
the superscript $\kappa$ for the fields, being now obvious that in the following we refer to elements of $\mathcal{S}^{\mathbb{C}}_{\kappa}$.}
$(\Phi_1,\Phi_2)_{\kappa}=-2i\omega_{\kappa}(\Phi^*_1,\Phi_2)$ which written explicitly reads
\begin{equation}
\label{kinnerp}
(\Phi_1,\Phi_2)_{\kappa}=\int\frac{d^4 p}{(2\pi)^3}\,\,\delta(\mathcal
{C}_{\kappa}(p))\,\, \mathrm{sign}{\left(e^{\frac{p_0}{\kappa}}-1\right)}
\,\,e^{\frac{3p_0}{\kappa}}\,\tilde{\Phi^*}_1(\dot{-}p)\tilde{\Phi}_2
(p)\, .
\end{equation}
in which $C_{\kappa}(p)$ is the deformed mass Casimir of the $\kappa$-Poincar\'e Hopf algebra. If in equation (\ref{kinnerp}) above we restrict to modes in the range $|\vec{p}|\in[\,0,\,\kappa\,)$ the factor $\mathrm{sign}{(e^{\frac{p_0}{\kappa}}-1)}$ can be written in the more familiar form $p_0/|p_0|$ and thus we have
\begin{equation}
\label{kinnerp2}
(\Phi_1,\Phi_2)_{\kappa}=\int\frac{d^4 p}{(2\pi)^3}\,\,\delta(\mathcal
{C}_{\kappa}(p))\,\,\frac{p_0}{|p_0|}
\,\,e^{\frac{3p_0}{\kappa}}\,\tilde{\Phi^*}_1(\dot{-}p)\tilde{\Phi}_2
(p)\, .
\end{equation}
Energy and momentum charges associated with $\kappa$-deformed translation symmetries for each $\Phi\in\mathcal{S}^{\mathbb{C}}_{\kappa}$ will be given by\footnote{
The definition of the action of translational symmetries on fields $P^{\kappa}_{\mu} \vartriangleright \Phi$ involves a choice of deformed differential calculus.  Such choice is not unique (see \cite{Freikono, FreiKoNo2} for examples of Noether analyses with a 5-d differential calculus). There appear to be no obstruction in adapting the results presented here to any choice of differential calculus.}
\begin{equation}
Q^{\kappa}_{\mu}=\frac{1}{2}(\Phi, P^{\kappa}_{\mu} \vartriangleright \Phi)_{\kappa}\, .
\end{equation}

\section{Quantum fields}
Now we move on to the quantization of the linear (non-interacting) field theory described in the previous sections.  We will first recall some basic facts about canonical quantization in the usual special relativistic framework keeping in mind the covariant phase space formalism used above.  
\subsection{Standard Hilbert space construction}
The ``one-particle" Hilbert space of a free quantum (scalar) field theory can be constructed starting from the complexified space of classical solutions of the equations of motion $\mathcal{S}^{\mathbb{C}}$ \cite{Wald:1995yp}.  In order to turn $\mathcal{S}^{\mathbb{C}}$ into a Hilbert space one needs to define an inner product.  The natural choice would be the conserved product (\ref{stinnerp}), that preserves the Poincar\'e symmetry structure, but it turns out that such product fails to be positive definite on all $\mathcal{S}^{\mathbb{C}}$, thus one has to restrict to a subspace of $\mathcal{S}^{\mathbb{C}}$ on which (\ref{stinnerp}) is positive definite.  One possible choice\footnote{The choice of a subspace of $\mathcal{S}^{\mathbb{C}}$ to be used as a Hilbert space is actually not unique.  Any subspace satisfying appropriate conditions (see e.g. \cite{Wald:1995yp}) would work.  Most interesting properties of quantum fields on curved space-time (and for non-inertial observers in flat space-time) can be traced back to this observation.} is to restrict to the subspace spanned by {\it positive frequency} solutions which we denote with $\mathcal{S}^{\mathbb{C}+}$.  The Hilbert space completion of $\mathcal{S}^{\mathbb{C}+}$ under (\ref{stinnerp}) is the one particle Hilbert space of the theory $\mathcal{H}$.  Using Fourier transforms it can be shown that $\mathcal{H}$ (and thus $\mathcal{S}^{\mathbb{C}+}$ as a vector space) is isomorphic to the Hilbert space of square integrable functions in momentum space on the positive mass shell.  One also has that $\mathcal{S}^{\mathbb{C}}$ is spanned by $\mathcal{H}$ and its complex conjugate $\bar{\mathcal{H}}$ and that for $\xi^+\in\mathcal{H},\, \xi^-\in\bar{\mathcal{H}}$ one has $(\xi^+,\xi^-)=0$.\\
From $\mathcal{H}$ one can construct the (symmetric) Fock space $\mathcal{F}_s(\mathcal{H})$ as follows:  denote $\mathcal{H}^n=\prod_{j=1}^n\otimes\mathcal{H}_j$ (with $\mathcal{H}^0=\mathbb{C}$); if $\sigma \in P_n$ is a permutation in the permutation group of $n$-elements and $\xi^+_1\otimes\cdots\otimes\xi^+_n$ is a basis of $\mathcal{H}^n$
\begin{equation}
\sigma(\xi^+_1\otimes\cdots\otimes\xi^+_n)=\xi^+_{\sigma(1)}\otimes\cdots\otimes
\xi^+_{\sigma(n)}
\end{equation}
one defines
\begin{equation}
S_n=\frac{1}{n!}\sum_{\sigma\in P_n}\sigma
\end{equation}
then the bosonic (symmetric) Fock space is given by
\begin{equation}
\mathcal{F}_s(\mathcal{H})=\bigoplus_{n=0}^{\infty}S_n\mathcal{H}^n\, .
\end{equation}\\
We denote a generic element $\Psi\in\mathcal{F}_s(\mathcal{H})$ as $\Psi=(a,a_1,a_2,\dots)$ with $a\in\mathbb{C}$, $a_1\in\mathcal{H}$, $a_2\in\mathcal{H}\otimes_s\mathcal{H}$ etc.  Observables are self-adjoint operators on $\mathcal{F}_s(\mathcal{H})$.  For each ``positive frequency" solution or ``one-particle" state $\xi\in\mathcal{H}$ one defines the {\it annihilation operator} $a(\bar{\xi}):\mathcal{F}_s(\mathcal{H})\rightarrow\mathcal{F}_s(\mathcal{H})$
\begin{equation}
a(\bar{\xi})\Psi=(\bar{\xi}\cdot a_1, \sqrt{2}\,\bar{\xi}\cdot a_2,\sqrt{3}\,\bar{\xi}\cdot a_3,\dots)
\end{equation}
where $\bar{\xi}\in\bar{\mathcal{H}}$ is the complex conjugate vector associated to $\xi$ and $\bar{\xi}\cdot a_n$ is the ``contraction", via inner product, of the first element of each tensor product of the symmetrized sum appearing in $a_n$ with $\bar{\xi}$.  The vacuum state $|\,0>\in\mathcal{F}_s$ is defined by 
\begin{equation}
a(\bar{\xi})|\,0>=0\,\,\forall \xi\in \mathcal{H}\, .
\end{equation}
The creation operator, adjoint to $a(\bar{\xi})$ is defined by
\begin{equation}
a^{\dagger}(\xi)\Psi=(0,\xi\, a, \sqrt{2}\,\xi\otimes_s a_1,\sqrt{3}\,\xi\otimes_s a_2,\dots)\,.
\end{equation}
If $\{\xi_i\}$ is an orthonormal basis of our (separable) Hilbert space $\mathcal{H}$ then the quantum field operator can be written as
\begin{equation}
\hat{\Phi}=\sum_i \left(\xi_i\,\, a(\bar{\xi}_i)+\bar{\xi}_i\,\, a^{\dagger}(\xi_i)\right)\, ;
\end{equation}
when such basis elements are (positive energy) plane waves $\phi^+_{\vec{p}}$ and we are restricting to a box of spatial size $L$ one obtains the familiar expression\footnote{From a rigorous mathematical point of view the ``field operator" is an operator valued distribution.  In the case of linear scalar field (i.e. a non interacting theory) our lack of rigor will be harmless.} 
\begin{equation}
\hat{\Phi}=\sum_{\vec{p}}\left(\phi^+_{\vec{p}}\,a_{\vec{p}}+\bar{\phi}^+_{\vec{p}}\,a^{\dagger}_{\vec{p}}\right)\, ,
\end{equation}
with $a^{\dagger}_{\vec{p}}=a_{-\vec{p}}$, as we are dealing with a real scalar field.\\
In the next section we will see how the above construction must be modified in the presence of deformations of the algebra of relativistic symmetries.  To this end it is useful to remind the reader how standard symmetries act on the algebra of scalar (quantum) fields equipped with the standard multiplication ``$\cdot$". As a symmetry of our theory $Y\in\mathcal{P}$ will act on the elements of $\mathcal{F}_s(\mathcal{H})$ as $Y\vartriangleright\Psi$ and, formally, on the operators defined on the Fock space through $Y\blacktriangleright a$.  We would like to establish a relation between these two actions.  First let's consider the action of a Poincar\'e group transfomation on an element of $\Psi\in\mathcal{F}_s(\mathcal{H})$:
\begin{equation}\label{poinctransf}
U(\Lambda,\varepsilon)(a \Psi)=a_{(\Lambda,\varepsilon)}\Psi_{(\Lambda,\varepsilon)}
\end{equation}
For an infinitesimal transformation $U(\Lambda,\varepsilon)(a \Psi)\simeq1-i\delta^cY_c$ (with $\delta^c$ infinitesimal parameters and $Y_c\in\mathcal{P}$) one has from (\ref{poinctransf}) that
\begin{equation}
(Y\blacktriangleright a)\Psi=(Y\vartriangleright a\Psi)-a(Y\vartriangleright\Psi)
\end{equation}
which can be rewritten formally as $Y\blacktriangleright a=[Y,a]$.  One can verify, using the above relation, that the following property must hold
\begin{equation}\label{condition}
Y\blacktriangleright(a\circ b)=\circ [\Delta Y\blacktriangleright (a\otimes b)]
=\sum(Y_{(1)}\blacktriangleright a)\circ(Y_{(2)}\blacktriangleright b)\, .
\end{equation}
where $\Delta Y$ is the trivial co-product of the standard Poincar\'e (Hopf) algebra (see (\ref{trivialcopro})), $\circ$ is the composition law for operators actiong on $\mathcal{F}_s(\mathcal{H})$ and in the last term we used Sweedler's notation for the co-product.
The equation above is telling us that there is a non-trivial relation between the co-product and the composition law for operators.  In particular if the co-product is deformed or ``twisted" then one has to modify $\circ$ to a deformed or ``star" product $\star$.  The need for a ``deformed" composition law for operators is well known for the case of $\theta$-twisted Poincar\'e algebra  in which only the co-product structure is deformed \cite{theta_quant.1, theta_quant.2, theta_quant.3, Joung:2007qv}.  As we will see in the next sections something similar emerges in the $\kappa$-Poincar\'e case in which the non-trivial co-algebra structure exhibits deformed (non-cocommutative) co-product as well as a deformed action of symmetry generators.\\
We can write down the explicit form of the action of the generators of translations and space-time rotations on the creation and annihilation operators, $a_{\vec{k}}, a^{\dagger}_{\vec{k}}$, from their action on one-particle states. The state $a^{\dagger}_{\vec{k}} |\,0>$ is, by construction, a positive energy eigenvector of the momentum operators with
\begin{equation}
P_0 \vartriangleright (a^{\dagger}_{\vec{k}} |\,0 >)=\omega(\vec{k})(a^{\dagger}_{\vec{k}} |\,0>)\,,\,\,\,\,\,\,
P_i \vartriangleright (a^{\dagger}_{\vec{k}} |\,0>)=k_i (a^{\dagger}_{\vec{k}} |\,0>)\, .
\end{equation}
From
\begin{equation}
P_{\mu}\vartriangleright (a^{\dagger}_k |\,0>)
= (P_{\mu}\blacktriangleright a^{\dagger}_k) |\,0>
+ a^{\dagger}_k  (P_{\mu}\vartriangleright |\,0>)
\end{equation}
and the invariance of the vacuum under translations we derive
\begin{equation}
P_0 \blacktriangleright a^{\dagger}_{\vec{k}} = \omega(\vec{k}) \,\, a^{\dagger}_{\vec{k}}\,,\,\,\,\,\,
P_i \blacktriangleright a^{\dagger}_{\vec{k}} = k_i\,\, a^{\dagger}_{\vec{k}}\, .
\end{equation}
For boosts and rotations one has
\begin{eqnarray}
M_{0j}\blacktriangleright a^{\dagger}_{\vec{k}} & = & i\, \omega(\vec{k}) \partial_{k_j} \,a^{\dagger}_{\vec{k}}\\
M_{jl}\blacktriangleright a^{\dagger}_{\vec{k}} & = & i \left( k_j\partial_{k_l}-k_l\partial_{k_j}\right) a^{\dagger}_{\vec{k}}\,\\
\end{eqnarray}
and analogous expressions, with appropriate sign changes, hold for annihilation operators $a_{\vec{k}}$.

\subsection{$\kappa$-one-particle Hilbert space}
In the following sections we propose an extension of the standard Fock space construction described above to the case when standard relativistic symmetries described in terms of the Poincar\'e (Lie) algebra are generalized by the $\kappa$-Poincar\'e deformed (Hopf) algebra of symmetries. We assume that the physical Hilbert space of our quantum field theory consists of a deformation of the standard bosonic Fock space $\mathcal{F}_{\kappa}(\mathcal{H})$ whose structure is, at this stage, unknown.\\  Our strategy will be to start from the space of solutions of the $\kappa$-Klein-Gordon equation from which, using the inner product (\ref{kinnerp}) (and implicitly the map (\ref{map})) described in Section II, we will define a $\kappa$-one-particle Hilbert space.  At this level the non-trivial Hopf algebra structure of the symmetries will impose a natural truncation to sub-planckian modes in the Hilbert space.  In the next section we will also show how the deformed co-algebra structure of $\kappa$-Poincar\'e will require a new concept of symmetrization when constructing ``$n$-particle" states leading to a non-trivial structure for the Fock space of the theory $\mathcal{F}_{\kappa}(\mathcal{H})$.  After defining a field operator we will apply our results to the calculation of the energy and momentum charges carried by one and two particle states of our theory and study their properties.\\ 
\newline
We start with the definition of our ``one-particle" Hilbert space.  To construct such Hilbert space we consider the inner product (\ref{kinnerp}) and look for a subspace of the space of complex solutions\footnote{In the following we will use $\mathcal{S}^{\kappa}$ instead of $\mathcal{S}^{\kappa}_{\mathbb{C}}$ to leave room for other indices.} $\mathcal{S}^{\kappa}$ on which such product is positive definite.  In
order to do so we give a closer look at the properties of our inner product
\begin{equation}
\nonumber
(\Phi_1,\Phi_2)_{\kappa}=\int\frac{d^4 p}{(2\pi)^3}\,\,\delta(\mathcal
{C}_{\kappa}(p))\,\, \mathrm{sign}{\left(e^{\frac{p_0}{\kappa}}-1\right)}
\,\,e^{\frac{3p_0}{\kappa}}\,\tilde{\Phi^*}_1(\dot{-}p)\tilde{\Phi}_2
(p)\, .
\end{equation}
One can easily verify that $ \mathrm{sign}(e^{\frac{p_0}{\kappa}}-1)$, when $p_0$ is put on-shell, behaves as
\begin{equation}
\label{thetasign}
\mathrm{sign}{\left(e^{\frac{p_0}{\kappa}}-1\right)}=\frac{p_0}{|p_0|}\,\theta(\kappa-|\vec{p}|)-\frac{p_0}{|p_0|} \,\theta(-
\kappa+|\vec{p}|)
\end{equation}
in which $\theta$ is the Heaviside function.  Now consider an orthonormal plane wave basis $\phi_{\vec{k}}=A\,e^{ikx}$ in which $A$ is a normalization factor\footnote{Without loss of generality we take $A$ to be real.}. Their $\kappa$-Fourier transform an its complex conjugate are given by
\begin{equation}
\tilde{\phi}_{\vec{k}}(p)= A\,(2\pi)^3\delta^{(3)}(\vec{k}-\vec{p})\,;\,\,\,\,\,(\tilde{\phi}_{\vec{k}}(p))^*=A\,(2\pi)^3\delta^{(3)}(\vec{p}-\vec{k})\,,
\end{equation}
and $\tilde{\phi^*}_k((\dot{-})p)=A\,(2\pi)^3\,e^{-3{\omega^+(|\vec{k}|)}/{\kappa}}\delta^{(3)}(\vec{p}-\vec{k})$.
Plugging these expressions in the formula for the inner product 
\begin{equation}
(\phi_{\vec{k}_1},\phi_{\vec{k}_2})_{\kappa}=(2\pi)^3 A^2 |\vec{k}_1|^{-1} \mathrm{sign}{\left(e^{\frac{k^0_1}{\kappa}}-1\right)}  \,\delta^{(3)}(\vec{k}_1-\vec{k}_2)\, 
\end{equation}
where $k^0_1$ is now on shell and from which we fix the normalization factor to $A=|\vec{k}|^{1/2}(2\pi)^{-3/2}$.  
As customary, we now put our fields ``in the box" so that the modes have finite spacing determined by the spatial extent of the box.\\
From (\ref{thetasign}) one can easily see that for modes on the ``positive energy" mass-shell ($\omega^+(\vec{k})$) with $|\vec{k}|>\kappa$ the inner product {\it is no longer positive definite}.  Indeed the energies associated with such modes become complex valued!  Thus, in order to construct our one-particle Hilbert space, we need to restrict to the subspace $\mathcal{S}^{\kappa +}_{|\vec{k}|\leq\kappa} \subset \mathcal{S}^{\kappa}$, the space of ``positive energy" solutions whose mode components are truncated at the Planck scale $\kappa$, on which the inner product is positive definite.  Our $\kappa$-one particle Hilbert space $\mathcal{H}_{\kappa}$ will be defined by the completion of $\mathcal{S}^{\kappa +}_{|\vec{k}|\leq\kappa} \subset \mathcal{S}^{\kappa}$ in the inner product (\ref{kinnerp}).\\
Through the $\kappa$-complex conjugation map described in Section II.B, whose definition combines standard complex conjugation and the antipode transformation, $\mathcal{S}^{\kappa+}_{|\vec{k}|\leq\kappa}$ is mapped into $\mathcal{S}^{\kappa-}$, the space of negative solutions, i.e. $\bar{\mathcal{S}}^{\kappa +}_{|\vec{k}|\leq\kappa}=\mathcal{S}_{}^{\kappa-}$.  Indeed, it can be easily seen that the antipode is a one-to-one and onto map only restricting to positive energy modes with $|\vec{k}|\leq\kappa$ or, equivalently, restricting to {\it purely real} energy modes.\\  
At this point, even though a general description of the Fock space of the theory $\mathcal{F}_{\kappa}(\mathcal{H})$ is still lacking, we can characterize the creation and annihilation operators relative to a specific basis in terms of their action on the vacuum state they single out in $\mathcal{F}_{\kappa}(\mathcal{H})$.  In particular consider the (normalized) plane wave basis $\{\phi^+_{\vec{p}}\}$ of  $\mathcal{S}^{\kappa +}_{|\vec{k}|\leq\kappa} \subset \mathcal{S}^{\kappa} $ described above. We define the annihilation operators $b_{\vec{p}}$ relative to $\{\phi^+_{\vec{p}}\}$ and a vacuum state $|\,0>\,\in\mathcal{F}_{\kappa}(\mathcal{H})$ such that
\begin{equation}
b_{\vec{p}}\,|\,0> \equiv (0,0,...)\,\,\, \forall \vec{p}
\, .
\end{equation}
The action of the creation operator $b^{\dagger}_{\vec{p}}$ on the vacuum will be defined by
\begin{equation}
b^{\dagger}_{\vec{p}} |\,0>=(0,\phi^+_{\vec{p}},0,...)\, .
\end{equation}
Our state $b^{\dagger}_{\vec{p}} |\,0>$ will be, according to the well defined action of momenta on plane waves (see \cite{Ale2}), a positive energy eigenvector of the momentum operators with
\begin{equation}
P_0 \vartriangleright (b^{\dagger}_{\vec{p}} |\,0>)=\omega^{+}(b^{\dagger}_{\vec{p}} |\,0>)\,;\,\,\,\,\,\, P_i \vartriangleright (b^{\dagger}_{\vec{p}} |\,0>)=p_i (b^{\dagger}_{\vec{p}} |\,0>)\, .
\end{equation}
in analogy with the standard case, we have
\begin{equation}
P_0 \blacktriangleright b^{\dagger}_{\vec{p}} = \omega^{+} \,\, b^{\dagger}_{\vec{p}}\,\,;\,\,\,\,\,\,P_i \blacktriangleright b^{\dagger}_{\vec{p}} = p_i\,\, b^{\dagger}_{\vec{p}}\, ,
\end{equation}
which again we write in a more compact form as
\begin{equation}
P_{\mu} \blacktriangleright b^{\dagger}_{\vec{p}}= p^{\mu}_{+} \,\, b^{\dagger}_{\vec{p}}\, .
\end{equation}
For the adjoint operators we have, according to the deformed complex conjugation defined in Section II.B 
\begin{equation}
P_{\mu}\blacktriangleright b_{\vec{p}} = (\dot{-} p^{\mu}_{+}) \,\, b_{\vec{p}}\, .
\end{equation}
Now that we have a candidate $\kappa$-one-particle Hilbert space we can start building multi-particle states and their associated Fock space.

\subsection{$\kappa$-Fock space and $n$-particle states}
In the case of standard scalar quantum fields, the Fock space $\mathcal{F}(\mathcal{H})$ built from the direct sum of tensor products of one-particle Hilbert spaces $\mathcal{H}$ is ``too large" and one has to resort to a symmetrization, related the specific type of particle statistics, which leads to the {\it bosonic} Fock space described in Section III.A.  The redundance in using $\mathcal{F}(\mathcal{H})$ as the Hilbert space of the theory comes from the fact that if we decide to label the states according to the number of particles and the eigenvalues of the four-momentum operator $P_{\mu}$, for an $n$-particle state there will be $n!$ different states in $\mathcal{F}(\mathcal{H})$ which have {\it the same} momentum eigenvalue.  One thus proceeds to symmetrize such states obtaining a single element belonging to the ``physical" Hilbert space $\mathcal{F}_s(\mathcal{H})$.  In the $\kappa$-deformed case one could try to proceed in an analogous way but such naive guess turns out to be inconsistent as the new symmetrized states, like e.g. the two-particle state\footnote{In the following we omit the superscript $+$ from $\phi^+_{\vec{p}}$ being at this point obvious that we are considering only ``positive energy" plane wave solutions.} $(0,0,1/\sqrt{2}(\phi_{\vec{p}}\otimes \phi_{\vec{q}}+\phi_{\vec{q}}\otimes \phi_{\vec{p}},0,...)$, will not be eigenstates of the operator $P_{\mu}$, in other words the standard ``flip" operator $\tau(\phi_{\vec{p}}\otimes \phi_{\vec{q}})=\phi_{\vec{q}}\otimes \phi_{\vec{p}}$ is no longer ``superselected" i.e. does not commute with all the observables. To overcome this difficulty, starting from the two-particle case, we will look for states which have the same momentum eigenvalues, extract the information regarding the type of non-trivial reshuffling of the momentum labels needed to go from one state to the other and then construct a $\kappa$-bosonic Fock space.\\
The two-mode tensor product state $\phi_{\vec{p}}\otimes \phi_{\vec{q}}$ has an associated four-momentum eigenvalue $p\dot{+}q$ due to the non-trivial co-product of the $P_{\mu}$'s .  We now look for momentum labels $\sigma^{\kappa}_{12}(\vec{p})$ and $\sigma^{\kappa}_{12}(\vec{q})$ such that $\phi_{\sigma^{\kappa}_{12}(\vec{q})}\otimes \phi_{\sigma^{\kappa}_{12}(\vec{p})}$ will carry the same four-momentum eigenvalue $p\dot{+}q$.  Here $\sigma_{12}$ indicates that the new momenta are associated to the only non-trivial element of the permutation group $S^2$ ($\sigma^{\kappa}_{0}$ being the identity) but {\it do not} correspond to the standard ``flip" of momentum labels. One can check that
\begin{eqnarray}
\sigma^{\kappa}_{12}(\vec{q})=\tilde{q} & = &(\tilde{q}_0;\,\vec{q}\,e^{-p_0/\kappa})\nonumber\\
\sigma^{\kappa}_{12}(\vec{p})=\tilde{p} & = &(\tilde{p}_0;\,\vec{p}\,e^{\tilde{q}_0/\kappa})
\label{bsymm}
\end{eqnarray}
with\footnote{Remember that, according to (\ref{kroots}), the time component of the four-momentum on the positive deformed mass-shell is given $\omega^+(\vec{p})=-\kappa\log(1-|\vec{p}|/\kappa)$} $\tilde{q}_0=\omega^+(\vec{q}\,e^{-\omega^+(\vec{p})/\kappa})$ and $\tilde{p}_0=
\omega^+(\vec{p}\,e^{\tilde{q}_0/\kappa})$ are such that $\tilde{q}\dot{+}\tilde{p}=p\dot{+}q$.  One also easily verifies that such choice is unique.\\
We can write down the action of two creation operators on the vacuum obtaining a $\kappa$-symmetrized two-particle state\footnote{From now on, in an effort to keep the notation clear, the different modes will be labeled in terms of the four-momentum $p$ instead of the spatial vector $\vec{p}$ and it will be understood that the fourmomenta are on shell.}
\begin{equation}
b^{\dagger}_{p}\star b^{\dagger}_{q} |\,0>=(0,0,1/\sqrt{2}(\phi_{p}\otimes \phi_{q}+\phi_{\tilde{q}}\otimes \phi_{\tilde{p}}),0,...)=
b^{\dagger}_{\tilde{q}}\star b^{\dagger}_{\tilde{p}} |\,0>\, .
\end{equation}
Notice how we introduced the {\it non-abelian} $\star$ composition of creation and annihilation operators, which reflects the non-cocommutativity of the coproduct, to stress that $b^{\dagger}_{p}\star b^{\dagger}_{q}$ and $b^{\dagger}_{q}\star b^{\dagger}_{p}$, acting on the vacuum, will create {\it two different} states in $\mathcal{F}^{\kappa}_s(\mathcal{H})$.  The action of an annihilation operator on the state above will be
\begin{equation}
b_{k}\star b^{\dagger}_{p}\star b^{\dagger}_{q} |\,0>= (0,(\delta_{\vec{k}\vec{p}}\,\,\phi_{q}+\delta_{\vec{k}\vec{\tilde{q}}}\,\,\phi_{\tilde{p}}),0,...)\, .
\end{equation}
A pictorial representation of how one gets from $p\dot{+}q$ to $\tilde{q}\dot{+}\tilde{p}$ will be helpful in generalizing the construction to $n$-mode states.  
We consider two modes of a single particle state as an an ordered couple $(\vec{p}_1;\,\vec{p}_2)$ and associate to each element of the couple a node of a graph.  Drawing an arrow going from node 1 to node 2 corresponding to the ordered couple $(\vec{p}_1;\,\vec{p}_2)$ will give the two-particle state labeled by the couple $(\vec{p}_1;\,\vec{p}_2\,e^{-p_1^0/\kappa})$ with momentum eigenvalue $p_1\dot{+}p_2$ (FIG. 1).
\begin{figure}
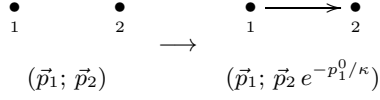

\[
\begin{array}{ccc}
\xy
(0,0)*{\bullet}+(0,-2.5)*{\scriptstyle{1}};
(14,0)*{\bullet}+(0,-2.5)*{\scriptstyle{2}};
\endxy 
& &
\xy
(0,0)*{\bullet}+(0,-2.5)*{\scriptstyle{1}};
{\ar (2,0)*{}; (12,0)*{}}; 
(14,0)*{\bullet}+(0,-2.5)*{\scriptstyle{2}};
\endxy\\
 &\,\,\,\,\longrightarrow\,\,\,\,& \\
(\vec{p}_1;\,\vec{p}_2)&
&(\vec{p}_1;\,\vec{p}_2\,e^{-p_1^0/\kappa})
\end{array}
\]
\caption{Two-particle state with momentum eigenvalue $p_1\dot{+}p_2$}
\end{figure}
Analogously drawing an arrow going from 2 to 1 gives the two-particle state $(\vec{p}_1\,e^{-p_2^0/\kappa};\,\vec{p}_2)$ with momentum eigenvalue $p_2\dot{+}p_1$ (FIG. 2).
\begin{figure}[htbp]
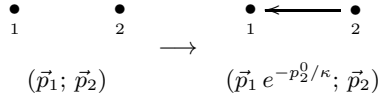

\begin{center}
\[
\begin{array}{ccc}
\xy
(0,0)*{\bullet}+(0,-2.5)*{\scriptstyle{1}};
(14,0)*{\bullet}+(0,-2.5)*{\scriptstyle{2}};
\endxy 
& &
\xy
(0,0)*{\bullet}+(0,-2.5)*{\scriptstyle{1}};
{\ar (12,0)*{}; (2,0)*{}}; 
(14,0)*{\bullet}+(0,-2.5)*{\scriptstyle{2}};
\endxy\\
 &\,\,\,\,\longrightarrow\,\,\,\,& \\
(\vec{p}_1;\,\vec{p}_2)&
&(\vec{p}_1\,e^{-p_2^0/\kappa};\,\vec{p}_2)
\end{array}
\]
\caption{Two-particle state with momentum eigenvalue $p_2\dot{+}p_1$}
\label{2p2}
\end{center}
\end{figure}
Notice how summing the elements of the resulting ordered couple in the first case reproduces the spatial component of the coproduct $p_1\dot{+}p_2$ and in the second case of $p_2\dot{+}p_1$.\\
``Retracting" an arrow going from node 1 to node 2 corresponding to the couple 
$(\vec{p}_1;\,\vec{p}_2)$ is the inverse move of the one described above and will give $(\vec{p}_1;\,\vec{p}_2\,e^{p_1^0/\kappa})$.  Retracting an arrow extending from 2 to 1 one goes from $(\vec{p}_1;\,\vec{p}_2)$ to $(\vec{p}_1\,e^{p_2^0/\kappa};\,\vec{p}_2)$ (FIG. 3).\\
\begin{figure}[htbp]
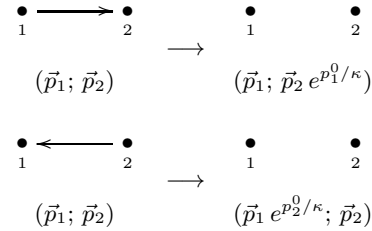

\begin{center}
\[
\begin{array}{ccc}
\xy
(0,0)*{\bullet}+(0,-2.5)*{\scriptstyle{1}};
{\ar (2,0)*{}; (12,0)*{}};
(14,0)*{\bullet}+(0,-2.5)*{\scriptstyle{2}};
\endxy 
& &
\xy
(0,0)*{\bullet}+(0,-2.5)*{\scriptstyle{1}};
 (14,0)*{\bullet}+(0,-2.5)*{\scriptstyle{2}};
\endxy\\
 &\,\,\,\,\longrightarrow\,\,\,\,& \\
(\vec{p}_1;\,\vec{p}_2)&
&(\vec{p}_1;\,\vec{p}_2\,e^{p_1^0/\kappa})\\
\\
\xy
(0,0)*{\bullet}+(0,-2.5)*{\scriptstyle{1}};
{\ar (12,0)*{}; (2,0)*{}}; 
(14,0)*{\bullet}+(0,-2.5)*{\scriptstyle{2}};
\endxy 
& &
\xy
(0,0)*{\bullet}+(0,-2.5)*{\scriptstyle{1}};
(14,0)*{\bullet}+(0,-2.5)*{\scriptstyle{2}};
\endxy\\
 &\,\,\,\,\longrightarrow\,\,\,\,& \\
(\vec{p}_1;\,\vec{p}_2)&
&(\vec{p}_1\,e^{p_2^0/\kappa};\,\vec{p}_2)
\end{array}
\]
\caption{``Undoing" two-particle states}
\label{default}
\end{center}
\end{figure}
In general an $n$-mode state graph will consist of $n$ nodes.  Each node will have $l$ arrows ending on it  with $0\leq l\leq n-1$ and each value of $l$ will occur only once in the graph.  We call $l$ the {\it weight} of the node.  Each one of the $n$-mode graphs has an associated  coproduct that we represent as a set of spatial momenta multiplied by the exponents determined by the coproduct.\\
Let's see how it is possible to obtain the correct symmetrized tensor product states using the pictorial representation just introduced.  Consider the simple example of a two-mode state with momentum eigenvalue $p_1\dot{+}p_2=(\omega^+_1(\vec{p}_1)+\omega^+_2(\vec{p}_2);\,\vec{p}_1+\vec{p}_2\,e^{-p_1^0/\kappa})$.  We can obtain such co-product summing the two entries of $(\vec{p}_1;\,\vec{p}_2\,e^{-p_1^0/\kappa})$ or $(\vec{p}_2\,e^{-p_1^0/\kappa};\,\vec{p}_1)$ (attach the first entry to node 1 and the second entry to node 2).  ``Undoing" the state in both cases gives the two components of the simmetryzed two-particle state given in (\ref{bsymm}) (see FIG. 4).\\  
\begin{figure}
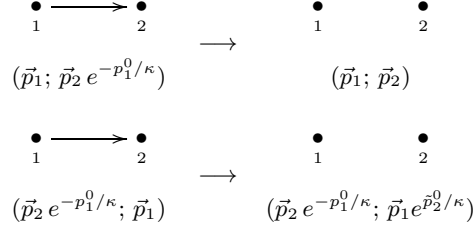

\label{2symm}
\[
\begin{array}{ccc}
\xy
(0,0)*{\bullet}+(0,-2.5)*{\scriptstyle{1}};
{\ar (2,0)*{}; (12,0)*{}}; 
(14,0)*{\bullet}+(0,-2.5)*{\scriptstyle{2}};
\endxy 
& &
\xy
(0,0)*{\bullet}+(0,-2.5)*{\scriptstyle{1}};
(14,0)*{\bullet}+(0,-2.5)*{\scriptstyle{2}};
\endxy\\
 &\,\,\,\,\longrightarrow\,\,\,\,& \\
(\vec{p}_1;\,\vec{p}_2\,e^{-p_1^0/\kappa})&
&(\vec{p}_1;\,\vec{p}_2)\\
&&\\
\xy
(0,0)*{\bullet}+(0,-2.5)*{\scriptstyle{1}};
{\ar (2,0)*{}; (12,0)*{}}; 
(14,0)*{\bullet}+(0,-2.5)*{\scriptstyle{2}};
\endxy 
& &
\xy
(0,0)*{\bullet}+(0,-2.5)*{\scriptstyle{1}};
(14,0)*{\bullet}+(0,-2.5)*{\scriptstyle{2}};
\endxy\\
 &\,\,\,\,\longrightarrow\,\,\,\,& \\
(\vec{p}_2\,e^{-p_1^0/\kappa};\,\vec{p}_1)&
&(\vec{p}_2\,e^{-p_1^0/\kappa};\,\vec{p}_1e^{\tilde{p}_2^0/\kappa})
\end{array}
\]
\caption{Obtaining the two states which contribute to the two-particle symmetrized state with momentum eigenvalue $p_1\dot{+}p_2$}
\end{figure}
This procedure can be easily generalized to any $n$-mode state: the $n!$ degeneracy of states will be given by attaching the $n$ addenda of the co-product to each of the $n$ nodes of the graph (total of $n!$ possible ways) and then retracting the arrows in the following sequence:\\
1. retract the arrow going from the node with weight 0 and the one with weight 1;\\
2. retract the 2 arrows going from the weight 0 nodes and the weight 2 node;\\
...\\
$n-1$.  retract the $n-1$ arrows going from the weight 0 nodes to the weight $n-1$ node.\\
Let us watch the above algorithm at work in the example of a 3-mode state.   We focus on the symmetrized 3-mode state with momentum eigenvalue $p_1\dot{+}p_2\dot{+}p_3=(\omega^+_1(\vec{p}_1)+\omega^+_2(\vec{p}_2)+\omega^+_3(\vec{p}_3);\,\vec{p}_1+\vec{p}_2\,e^{-p_1^0/\kappa}+\vec{p}_3\,e^{-p_1^0/\kappa-p_2^0/\kappa})$ represented by the graph in FIG.5.  We are looking for the $3!$ tensor product states belonging to $\mathcal{H}_{\kappa}\otimes\mathcal{H}_{\kappa}\otimes\mathcal{H}_{\kappa}$ with the same momentum eigenvalue $p_1\dot{+}p_2\dot{+}p_3$.  These states will be obtained by attaching each one of the elements of the triple $(\vec{p}_1,\vec{p}_2\,e^{-p_1^0/\kappa},\vec{p}_3\,e^{-p_1^0/\kappa-p_2^0/\kappa})$ to the nodes of the graph in FIG. 5
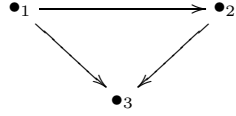
\begin{figure} \centering 
\[ 
\xymatrix{ 
{\bullet}_{\scriptstyle{1}}
\ar[rr] 
\ar[dr]
&& {\bullet}_{\scriptstyle{2}}
\ar[dl] \\ 
& {\bullet}_{\scriptstyle{3}}}
\]
\caption{The 3-particle state with momentum eigenvalue $p_1\dot{+}p_2\dot{+}p_3$.} 
\end{figure} 
in all the $3!$ possible ways, which correspond to the permutations of the elements of the triple, and then ``undoing" the arrows according to the algorithm above.  We can label each of the $3!$ addenda of the $\kappa$-symmetrized state with elements of the permutation group of three elements.  For example $\sigma^{\kappa}_{12}$ will stand for the element obtained starting from the permutation which exchanges the first and the second term in the triple, $\sigma^{\kappa}_{123}$  to the one which exchanges the first and the second terms and then the second and the third etc.   The triple obtained from the identity element $\sigma^{\kappa}_0$ (i.e. $\vec{p}_1$ attached to node 1, $\vec{p}_2\,e^{-p_1^0/\kappa}$ attached to node 2 etc.) would be $(\vec{p}_1,\vec{p}_2,\vec{p}_3)$ and is the analogous of the first couple appearing on the right-hand side of FIG. 4.
Putting $\alpha_i=\frac{|\vec{p}_i|}{\kappa}$ one can write down the whole set of tensor products which summed will give the symmetrized 3-particle state
\begin{align}
\sigma^{\kappa}_0=&\left(\vec{p}_1,\vec{p}_2,\vec{p}_3\right)\\
\sigma_{23}^{\kappa}=&\left(\vec{p}_1,(1-\alpha_2)\vec{p}_3,(1-\alpha_3(1-\alpha_2))^{-1}\vec{p}_2\right)\\
\sigma_{12}^{\kappa}=&\left((1-\alpha_1)\vec{p}_2,(1-\alpha_2(1-\alpha_1))^{-1}\vec{p}_1,\vec{p}_3\right)\\
\sigma_{123}^{\kappa}=&\Big((1-\alpha_1)\vec{p}_2,(1-\alpha_2(1-\alpha_1))^{-1}(1-\alpha_1)(1-\alpha_2)\vec{p}_3,  \nonumber \\  &(1-\alpha_2(1-\alpha_1))^{-1}(1-\alpha_3(1-\alpha_2(1-\alpha_1))^{-1}(1-\alpha_1)(1-\alpha_2))^{-1}\vec{p}_1\Big)\\
\sigma_{13}^{\kappa}=&\Big((1-\alpha_1)(1-\alpha_2)\vec{p}_3,(1-\alpha_3(1-\alpha_1)(1-\alpha_2))^{-1} (1-\alpha_1)\vec{p}_2,  \nonumber \\  & (1-\alpha_3(1-\alpha_1)(1-\alpha_2))^{-1}
(1-\alpha_2(1-\alpha_3((1-\alpha_1)(1-\alpha_2))^{-1}(1-\alpha_1))^{-1}
\vec{p}_1\Big)\\
\sigma_{132}^{\kappa}=&\Big((1-\alpha_1)(1-\alpha_2)\vec{p}_3,(1-\alpha_3(1-\alpha_1)(1-\alpha_2))^{-1} \vec{p}_1,  \nonumber \\  & (1-\alpha_3(1-\alpha_1)(1-\alpha_2))^{-1}
(1-\alpha_1(1-\alpha_3(1-\alpha_1)(1-\alpha_2))^{-1})^{-1}
(1-\alpha_1)\vec{p}_2\Big)\, .
\end{align}
To clarify the above formulae note that using the same notation the elements of the symmetrized two-mode state would have been written as
\begin{align}
\sigma^{\kappa}_0=&\left(\vec{p}_1,\vec{p}_2\right)\\
\sigma_{12}^{\kappa}=&\left((1-\alpha_1)\vec{p}_2,(1-\alpha_2(1-\alpha_1))^{-1}\vec{p}_1\right)
\end{align}
The 3-mode state with momentum eigenvalue $p_1\dot{+}p_2\dot{+}p_3$ can finally be written as
\begin{equation}
\label{3ps}
b^{\dagger}_{p_1}\star b^{\dagger}_{p_2}\star b^{\dagger}_{p_3} |\,0>=\left(0,0,0,1/\sqrt{6}
\left(\sigma^{\kappa}_0+\sigma_{23}^{\kappa}+\sigma_{12}^{\kappa}+\sigma_{123}^{\kappa}+
\sigma_{13}^{\kappa}+\sigma_{132}^{\kappa}\right),0,...\right)\, ,
\end{equation}
where each $\sigma^{\kappa}$ corresponds to an ordered triple above which in turn corresponds to a tensor product of plane wave solutions with given momentum labels $(\vec{p}_1,\vec{p}_2,\vec{p}_3)\rightarrow\phi_{p_1}\otimes\phi_{p_2}\otimes\phi_{p_3}$.
Applying an annihilation operator we have
\begin{align}
b_{p}\star b^{\dagger}_{p_1}\star b^{\dagger}_{p_2}\star b^{\dagger}_{p_3} |\,0>=&\Big(0,0,1/\sqrt{2}\Big[\delta_{\vec{p}\,,\vec{p}_1}\left(\bar{\sigma}^{\kappa}_0+\bar{\sigma}_{23}^{\kappa}\right)+
\delta_{\vec{p}\,,(1-\alpha_1)\vec{p}_2}\left(\bar{\sigma}_{12}^{\kappa}+\bar{\sigma}_{123}^{\kappa}\right)+\nonumber \\+&\delta_{\vec{p}\,,(1-\alpha_1)(1-\alpha_2)\vec{p}_3}\left(\bar{\sigma}_{13}^{\kappa}+\bar{\sigma}_{132}^{\kappa}\right)\Big],...\Big)\, .
\end{align}
where $\bar{\sigma}_{...}^{\kappa}$ stands for one of the addenda in (\ref{3ps}) with the first factor in the tensor product removed.  We have written down these states explicitly as we will need such expressions in order to calculate the Noether charges carried by multi-particle states in the following section.\\  Having a general prescription for the construction of $n$-particle, $\kappa$-symmetrized, states from the one-particle Hilbert space $\mathcal{H}_{\kappa}$, we can define the deformed Fock space of the theory to be 
\begin{equation}
\mathcal{F}_{\kappa}(\mathcal{H})=\bigoplus_{n=0}^{\infty}S^{\kappa}_n\mathcal{H}_{\kappa}^n\, 
\end{equation}
where
\begin{equation}
S^{\kappa}_n=\frac{1}{n!}\sum_{\sigma\in P_n}\sigma^{\kappa}
\end{equation}
and to each permutation $\sigma\in P_n$ is associated a non trivial reshuffling of modes $\sigma^{\kappa}$
\begin{equation}
\sigma^{\kappa}(\phi_{k_1}\otimes\cdots\otimes\phi_{k_n})=\phi_{\sigma^{\kappa}(k_1)}\otimes\cdots\otimes
\phi_{\sigma^{\kappa}(k_n)}
\end{equation}
according to the rules described above.  Now that we have the full Hilbert space of our theory we can write down the deformed field operator acting on it
\begin{equation}
\hat{\Phi}=
\sum_{\vec{p},\,|\vec{p}|\leq\kappa}
\left(\phi_{p}b_{\vec{p}}+
\bar{\phi}_{p}b^{\dagger}_{\vec{p}}
\right)\, .
\end{equation}
and the $\kappa$-deformed counterpart of the reality condition for creation and annihilation operators reads
\begin{equation}
b^{\dagger}_{\vec{p}}=b_{\dot{-}\vec{p}}
\end{equation}
according to the notation for the antipode of modes on-shell introduced in Section II.B.

\subsection{Energy and momentum of quantum states}
As an application of the formalism introduced we now calculate the energy and momentum charges carried by  different states of our theory.  The Noether charges described in Section II.B
\begin{equation}
Q_{\mu}=\frac{1}{2}(\Phi, P_{\mu} \vartriangleright \Phi)_{\kappa}\, 
\end{equation}
in the quantum context  will become observables, i.e. operators on $\mathcal{F}_{\kappa}(\mathcal{H})$
\begin{equation}
\hat{Q}_{\mu}=\frac{1}{2}(\hat{\Phi}, P_{\mu} \vartriangleright \hat{\Phi})_{\kappa}\, ,
\end{equation}
whose expression depends on the explicit form of the field operator $\hat{\Phi}$ and where $P_{\mu} \vartriangleright \hat{\Phi}$ indicates that the generators of translations act on the ``function" coefficients of $\hat{\Phi}$.  Using the ortogonality properties
\begin{equation}
\left(\phi_{p}\, ,\bar{\phi}_{q}\right)=0\,,\,\,\,
\left(\phi_{p}\, ,\phi_{q}\right)=\delta_{\vec{p}\vec{q}}\,,\,\,\,
\left(\bar{\phi}_{p}\, ,\bar{\phi}_{q}\right)=-\,e^{-3\omega^+(\vec{p})/\kappa}\delta_{\vec{p}\vec{q}}
\end{equation}
and also
\begin{equation}
\left(\phi_{p}b_{\vec{p}}\, ,\phi_{q}b_{\vec{q}}\right)=\delta_{\vec{p}\vec{q}}\,\,
b^{\dagger}_{\vec{p}}\star b_{\vec{q}}\,,\,\,\,\,
\left(\bar{\phi}_{p}b^{\dagger}_{\vec{p}}\, ,\bar{\phi}_{q}b^{\dagger}_{\vec{q}}\right)=-
\,e^{-3\omega^+(\vec{p})/\kappa}\delta_{\vec{p}\vec{q}}\,\,b_{\vec{p}}\star
b^{\dagger}_{\vec{q}}
\end{equation}
we can write an explicit expression for the operators $\hat{Q}_{\mu}$  
\begin{equation}
\hat{Q}_{\mu}=\frac{1}{2}\sum_{\vec{p},\,|\vec{p}|\leq\kappa}\left[p_{\mu}\,b^{\dagger}_{\vec{p}}\star b_{\vec{p}}-\,e^{-3\omega^+(\vec{p})/\kappa}(\dot{-}p_{\mu})b_{\vec{p}}\star b^{\dagger}_{\vec{p}}\right]\,. 
\end{equation}
We start by calculating the energy-momentum charges carried by the one-particle state  
\begin{equation}
|\, p>=b^{\dagger}_p|\, 0>=\left( 0,\, \phi_{p},\, 0...\right)\, .
\end{equation}
Using the the explicit form of the operator $\hat{Q}_{\mu}$ given above one has
\begin{equation}
\label{1pcharge}
<\, p\, | \hat{Q}_{\mu} |\, p>=\frac{1}{2}(p^+_{\mu}- p^-_{\mu})-\frac{1}{2}\left(\sum_{\vec{k},\,|\vec{k}|\leq\kappa} e^{-3\frac{\omega^+(\vec{k})}{\kappa}}(\dot{-}k^+_{\mu})\right)
\end{equation}
where
\begin{eqnarray}
p_{\mu}^+&=&(\omega^+(\vec{p}), \vec{p})\nonumber\\
p_{\mu}^-&=&(\omega^-(\vec{p}), -\vec{p})\, .
\end{eqnarray}
The last term on the right hand side of (\ref{1pcharge}) is the analogous of the standard ``vacuum energy" and, due to the presence of the cut-off $\kappa$, turns out to be finite.  Once the vacuum contribution has been subtracted the one-particle energy momentum reads 
\begin{eqnarray}
<\, p\, | \hat{Q}_{\mu} |\, p>=\frac{1}{2}\left(p_{\mu}^+-p_{\mu}^- \right)\, .
\end{eqnarray}
We can write down an explicit form of the one-particle energy and momentum charges 
\begin{eqnarray}
Q_0^{(1-p)}&=&1/2(p_0^+-p_0^-) \nonumber\\
\vec{Q}^{(1-p)}&=&\vec{p}\, .
\end{eqnarray}
It is easily checked that they obey the dispersion relation
\begin{eqnarray}
|\vec{Q}|=\kappa \tanh\left(\frac{Q_0}{\kappa}\right)\, .
\end{eqnarray}
Notice how, as one would expect in the presence of a cut-off, the modulus of spatial momentum charge is bounded by $\kappa$, while the energy charge carried by the state goes to infinity when the modulus of the spatial momentum becomes planckian.\\
We now calculate the same charges for a two particle state.  As showed in the previous section, given two modes with momenta labels $p$ and $q$, we have two 2-particle states labeled by different values of the total fourmomentum:
 \begin{equation}
b^{\dagger}_{p}\star b^{\dagger}_{q}|\,0>=|\,p\dot{+}q>\,\,\,\,\,\,\,\,\,\, b^{\dagger}_{q}\star b^{\dagger}_{p}|\,0>=|q\dot{+}p>\, .
\end{equation}
Using the expressions for the action of four creation and annihilation operators acting on vacuum given in the previous subsection one obtains
\begin{equation}
<p\dot{+}q\,|\hat{Q}_{\mu}|\,p\dot{+}q>=Q^{Vac}_{\mu}+\left[ \frac{1}{2}\left( p^+_{\mu}+\tilde{q}^+_{\mu} \right)- \frac{1}{2}\left( p^-_{\mu}+\tilde{q}^-_{\mu} \right)  \right]
\end{equation}
in which 
\begin{eqnarray}
\tilde{q}_{\mu}^+=(\omega^+(\vec{q}e^{-p_0^+/\kappa}),\vec{q}e^{-p_0^+/\kappa})\nonumber\\
\tilde{q}_{\mu}^-=(\omega^-(\vec{q}e^{-p_0^+/\kappa}),-\vec{q}e^{-p_0^+/\kappa})
\end{eqnarray}
and
\begin{equation}
Q^{Vac}_{\mu}=-\frac{1}{2}\left(\sum_{\vec{k},\,|\vec{k}|\leq\kappa} e^{-3\frac{\omega^+(\vec{k})}{\kappa}}(\dot{-}k^+_{\mu})\right)
\end{equation}
The conserved charges for the two particle state $|\,p\dot{+}q> $ are then 
\begin{eqnarray}
Q_0^{(2-p)}&=&1/2(p_0^+-p^-_0)+1/2(\tilde{q}_0^+-\tilde{q}^-_0)\nonumber\\
Q_i^{(2-p)}&=&p_i+e^{-p_0^+ /\kappa}q_i\, .
\end{eqnarray}
Note how such charges are different from the corresponding ones carried by the state $|\,q\dot{+}p>$.  In fact such state has an associated energy $Q_0=1/2(q_0^+-q^-_0)+1/2(\tilde{p}_0^+-\tilde{p}^-_0)$, with $\tilde{p}=(\omega(\vec{p}e^{-q_0^+/\kappa}), \vec{p}e^{-q_0^+ /\kappa})$, and spatial momentum components $Q_i=q_i+e^{-q_0^+/\kappa}p_i$.  This indicates that the two states
$|\,p\dot{+}q>$ and $|q\dot{+}p>$ are truly different physical states.\\  Using the technology developed so far one could go on and calculate the energy-momentum charges for states with an arbitrary number of particles.  For our illustrative purposes here we limit to write explicit formulae up to the two-particle case.

\section{Comparison with previous analyses}
Before concluding we review and discuss some recent efforts in constructing quantum fields with $\kappa$-Poincar\'e symmetries\footnote{For the purpose of clarity, in this Section, we conform to
the notation used respectively in each work discussed.}.\\
In \cite{Lu} the authors attempt a quantization of a real scalar field on $\kappa$-Minkowski space-time with symmetries described by the $\kappa$-Poincar\'e algebra in the so-called ``symmetric basis".  Such basis is related to the standard bicrossproduct basis via the mapping $P_i\rightarrow{}P_i e^{\frac{P_0}{2\kappa}}$.  The authors write the following Fourier expansion of the field
\begin{eqnarray}
\,\phi(\hat{x})&=&\frac{1}{(2\pi)^{3/2}}\int d^4p \,
A(p_0,\vec{p})\,\delta\,\left(C_2^{\kappa}-M^2\right)
:e^{ip_{\mu}\hat{x}^{\mu}}: \nonumber
\end{eqnarray}
in which $M$ is a mass parameter and $C_2^{\kappa}$ represents the mass Casimir of the $\kappa$-Poincar\'e Hopf-algebra in the specific basis considered.  The Fourier coefficients of the
fields are on-shell according to the roots of the mass Casimir $p_0^{\pm}=\pm \omega_{\kappa}(\vec{p})$ and they are such that $\left(A(\pm\omega_{\kappa}(\vec{p}), \vec{p})
\right)^\dagger=A(\mp\omega_{\kappa}(\vec{p}), -\vec{p})$ holds.
Classical fields are promoted to quantum fields in \cite{Lu} by defining creation and annihilation operators given by
\begin{equation}
a_{\kappa}(\omega_{\kappa}(\vec{p}), \vec{p})=
C(\vec{p})\,A(\omega_{\kappa}(\vec{p}), \vec{p})\qquad
a_{\kappa}^{\dagger}(\omega_{\kappa}(\vec{p}), \vec{p})=
C(\vec{p})\,A(-\omega_{\kappa}(\vec{p}), -\vec{p}) \nonumber
\end{equation}
where the factors $C(\vec{p})$ result from the expansion of the delta function which sets the fields on-shell.\\
Moving to the multi-particle sector of the theory the authors adopt a strategy used in various works on
field theories with $\theta$-twisted Poincar\'e symmetries \cite{theta_quant.1, theta_quant.2, theta_quant.3}.  Namely they look for a composition rule $\circ$ for annihilation (and creation) operators which is consistent with the non-trivial co-product of $\kappa$-Poincar\'e in the symmetric basis
\begin{equation}
\Delta(P_0)=P_0\otimes1+1\otimes P_0 \qquad \Delta(P_i)=P_i\otimes
e^{\frac{P_0}{2\kappa}}+e^{-\frac{P_0}{2\kappa}}\otimes P_i \nonumber\,.
\end{equation}
i.e. such that the action of the $P_{\mu}$ generators on the annihilation operators is
\begin{eqnarray}
&P_{\mu}\,\triangleright \,a_{\kappa}\left(p_0, \vec{p}
\right)\,=\,p_{\mu} \,a_{\kappa}\left(p_0, \vec{p} \right)\,,\qquad
P_{\mu}\,\triangleright \,a^{\dagger}_{\kappa}\left(p_0, \vec{p}
\right)\,=\,-p_{\mu}\, a^{\dagger}_{\kappa}\left(p_0, \vec{p}
\right)\nonumber\\
&P_{\mu}\,\triangleright \,\left(a_{\kappa}\left(p_0, \vec{p}
\right)\,\circ a_{\kappa}\left(q_0, \vec{q} \right)\right)\,=\, \left(
\Delta^{(1)}(P_{\mu})\, \triangleright a_{\kappa}\left(p_0, \vec{p}
\right) \right)\, \left( \Delta^{(2)}(P_{\mu})\, \triangleright
a_{\kappa}\left(q_0, \vec{q} \right) \right) \nonumber\, .
\end{eqnarray}
The deformed composition rule derived by the authors is
\begin{equation}
a_{\kappa}(p)\, \circ \,
a_{\kappa}(q)\,:=\,a_{\kappa}\left(p_0,e^{-\frac{q_0}{2\kappa}}
\vec{p} \right) a_{\kappa}\left(q_0, e^{\frac{p_0}{2\kappa}}\vec{q}
\right)    \nonumber
\end{equation}
in which $p_0=\omega_{\kappa}(\vec{p})$, $q_0=\omega_{\kappa}(\vec{q})$.  Using this deformed ``multiplication" of operators the appropriate commutators $[\,,\,]_{\circ}$ are then written down and the authors observe that they simply reproduce the standard algebra of creation and annihilation operators.  This conclusion is not surprising since the choice of deformed composition of operators above tries to get rid of the non-trivial structure of the co-product.   However a major problem arises if  one notices that creation and annihilation operators, acting according to the deformed composition law given above, {\it do not create or destroy particles which are on-shell}\footnote{On the other hand it easily checked that such basic requirement is verified in the type of $\kappa$-symmetrization we propose in Section III.C.}, in fact $M^2=C^\kappa_2(k_0,\,\vec{k})\equiv \left(2\kappa\sinh\left( \frac{k_0}{2\kappa}\right)  \right)^2-\vec{k}^2$ is not satisfied, for example, for the mode $(k_0,\,\vec{k})\,=\,(p_0,\,e^{-\frac{q_0}{2\kappa}} \vec{p})$ belonging to a 2-particle state.  This undesirable feature represents, in our opinion, a major flaw of the ``quantization" of fields proposed in \cite{Lu}. 
\\
Two recent works in which quantization of $\kappa$-fields is discussed, in the context of an analysis of blackbody radiation and Unruh effect in $\kappa$-Minkowski space-time, are \cite{KRY1} and \cite{KRY2}.  There the authors define a real scalar field as the Fourier decomposition of plane waves which are taken to be the ``time to the left" ordered elements of the non-commutative coordinates $\hat{x}$, \i.e. $:e^{-i p\cdot \hat
{x}}:=e^{-i p^0 \hat{x}^0} e^{i \textbf{p}\cdot
\hat{\textbf{x}}}$. The basis of the $\kappa$-Poincar\'e Hopf algebra associated with such a normal ordering is related to the standard bicrossproduct basis by the redefinition  $P_i\rightarrow P_i
e^{\frac{P_0}{\kappa}}$ and by setting $\kappa\rightarrow{}-\kappa$. This basis is called by the authors ``asymmetric". In \cite{KRY2} the following field
expansion is used
\begin{eqnarray}
\phi(x)=\int_p e^{-i p\cdot x} f(p)\delta\left( \mathcal{E}(p) \right)
\,,\qquad \mathcal{E}(p)+m^2=M_{\kappa}^2(p)\left(
1+\frac{M_{\kappa}^2(p)}{4\kappa^2}  \right) \nonumber
\end{eqnarray}
where $M_{\kappa}^2(p)$ is the mass Casimir relative to the specific basis of the $\kappa$-Poincar\'e algebra considered. The deformed action for a scalar field used in \cite{KRY1} and \cite{KRY2} is
\begin{eqnarray}
S=\int_{\tilde{p}} \varphi^\dagger(p)\Delta_F^{-1}(p) \varphi(p)\,,
\qquad
\Delta_F^{-1}(p)=\mathcal{E}(p)+i\epsilon   \nonumber
\end{eqnarray}
where $\Delta_F^{-1}(p)$ is denoted as ``the Feynman propagator", $\tilde{p}$ is the antipode and $p\rightarrow-\tilde{p}$ is called the conjugate transformation.\\
The authors of \cite{KRY1} and \cite{KRY2} then perform a field quantization promoting the classical function $\Delta_F^{-1}(p)$ to the Feynman propagator of the quantum theory in momentum space.  Namely, as the authors write, ``it is technically safe to assume" that
\begin{equation}
i\Delta_F(p)\Big|_{on-shell}=\left<  \Phi(p) \Phi(p)^\dagger \right>
\nonumber
\end{equation}
The expansion of the quantum field is obtained imposing consistency with the latter relation. The authors then find an algebra of creation and annihilation operators that differs from the standard
one by a multiplicative factor $E(\textbf{q})$, i.e.
\begin{equation}
[\,a(\textbf{p}),a^\dagger(\textbf{q})\,]\,=\,2E(\textbf{p})\,(2\pi)^3
\,\delta^3(\textbf{p}\,-\,\textbf{q})\,,
\qquad [\,b(\textbf{p}),b^\dagger(\textbf{q})\,]\,=\,2E(\textbf{p})\,
(2\pi)^3\,\delta^3(\textbf{p}\,-\,\textbf{q})\,.\nonumber
\end{equation}
The authors repeat their analysis for the same basis of generators of $\kappa$-Poincar\'e considered in \cite{Lu}. The difference now emerges at the level of the coproduct for the translation generators  and at the level of the mass Casimir. The algebra of creation and annihilation operators obtained in this case \cite{KRY2} differs from the one relative to the ``asymmetric" basis by the replacement $E
(\textbf{p})\rightarrow{}D(\textbf{p})$. The possibility of re-absorbing both factors $E(\textbf{p})$ and $D(\textbf{p})$ by a suitable renormalization of the creation and annihilation operators is not discussed.\\
The authors of \cite{KRY1} and \cite{KRY2} essentially follow the quantization procedure, first proposed in \cite{KoLuMa}, which relies upon the assumption that the function $\Delta_F^{-1}(p)$
can be promoted to the Feynman propagator of the quantum theory.  The inconsistencies of such approach, which are particularly severe in the multi-particle sector, were already pointed out by the same authors of \cite{KoLuMa}.  It is not clear to us how such problems can be resolved in the case of \cite{KRY1} and \cite{KRY2}.  Moreover the analyses reported in \cite{KRY1} and \cite{KRY2} lack an explicit construction of the Hilbert space of the theory where, as we showed in the present work, the non-trivial co-algebra sector of the $\kappa$-Poincar\'e Hopf-algebra plays a crucial role.

\section{Conclusions}
We have constructed the Fock space for a free massless scalar field in the presence of deformed symmetries described by the $\kappa$-Poincar\'e Hopf algebra. On such space we have defined the basic observable of the theory, the field operator, and calculated the energy-momentum charges carried by one and two-particle states\\
The two main features of the space of quantum states we presented are the existence of planckian cut-off for the field modes in the one-particle sector and the need of a non-trivial bosonization in the construction of the Fock space of the theory.  Moreover the ``vacuum energy", which in usual quantum field theory is a divergent quantity, due to the presence of the cut-off $\kappa$ is now finite and, in our particular framework, energy and momentum charges of a single particle state obey a deformed dispersion relation.\\
To our knowledge this is the first time that a Hilbert space for quantum fields enjoying $\kappa$-Poincar\'e symmetries has been constructed and, remarkably, it provides an example of a quantum field theory in which an ultraviolet cut-off pacifically coexists with its relativistic symmetries.  The non-trivial behavior of the theory in the multi-particle sector is not surprising since, for example, it is well known that in the case of $\theta$-Poincar\'e Hopf algebra symmetries the usual flip operator must be ``twisted" thus leading to a deformed particle statistics \cite{theta_quant.1, theta_quant.2, theta_quant.3}.  However the type of non-trivial bosonization we found, characterized by an ``entanglement" of multiparticle modes, radically differs from the more popular framework of ``twisted" statistics and it is a challenge for future studies to investigate the physical consequences of such peculiar behavior.  Finally it is important to notice the emphasis of our approach to quantization on the definition of an inner product, creation and annihilation operators and field modes.  This emphasis makes the framework developed in the present paper tailored for generalizations to different choices of  ``time translation" Killing vectors, which define ``positive energy" modes for the field, and thus for the study of Unruh-like phenomena in a $\kappa$-deformed setting.  Such analysis will be the subject of a forthcoming work. 

\begin{acknowledgments}
We are very grateful to Giovanni Amelino-Camelia for stimulating discussions during the development of the present work.  We are indebted to Simone Severini for valuable suggestions regarding the ``pictorial" aspects of the $\kappa$-symmetrization.  We would also like to thank Bianca Dittrich, Laurent Freidel and Jurek Kowalski-Glikman for helpful remarks.  AM would like to thank Perimeter Institute for hospitality while part of this work was being carried out.\\
Research at Perimeter Institute for Theoretical Physics is supported in
part by the Government of Canada through NSERC and by the Province of
Ontario through MRI.
\end{acknowledgments}

\end{document}